\documentclass[useAMS,usenatbib]{mn2e}
\usepackage{natbib}
\usepackage{txfonts}
\bibpunct{(}{)}{;}{a}{}{,}
\usepackage[english]{babel}
\usepackage{ulem}

\usepackage{ifpdf}
\ifpdf 
\usepackage[pdftex]{graphicx}
\usepackage{epstopdf}
\else 
\usepackage{graphicx}
\fi

\newcommand{\be}{\begin{eqnarray}}
\newcommand{\ee}{\end{eqnarray}}

\def\figsize{5.5in}

\usepackage{colordvi}
\usepackage[usenames,dvipsnames]{xcolor}

\RequirePackage[colorlinks=true
,urlcolor=blue
,anchorcolor=blue
,citecolor=blue
,filecolor=blue
,linkcolor=blue
,menucolor=blue
,pagecolor=blue
,linktocpage=true
,pdfproducer=medialab
]{hyperref}

\def\msun{\mathrm{M}_\odot}

\begin{document}


\title{Long term evolution  of  an interacting binary system}

\author[Shaviv et al.]{Giora Shaviv$^1$,
Irit Idan$^1$ and Nir J. Shaviv$^{2,3}$\\
\noindent
$^1${Dept. of Physics, Israel Institute of Technology, Haifa 32000, Israel}\\
\noindent
$^2${The Racah Institute of physics, The Hebrew University of Jerusalem, Jerusalem 91904, Israel} \\
\noindent
$^3${School of Natural Sciences, Institute for Advanced Study, Princeton NJ, 08540, USA}}

\pagerange{\pageref{firstpage}--\pageref{lastpage}} \pubyear{2014}

\maketitle

\begin{abstract}
We describe a new code to simulate the stellar evolution of a close interacting binary system. It is then used to calculate the evolution of a classical nova system composed of a $1.25~\msun$ Main-Sequence (MS) star and a $1.0~\msun$ white dwarf (WD) companion. The system begins as a well separated non-interacting binary system. Initially, the two stars evolve  independently of each other. However, Roche lobe overflow begins as the MS star expands on its way to become a Red Giant. We follow the mass accreted onto the WD and the ensuing nuclear runaways for several thousand flashes.
The main finding is that the Roche-Lobe mass transfer rate is modulated by oscillations in the MS star, with a period that is somewhat shorter than the thermal time scale of the star. This periodically modulates the rate of thermonuclear flashes on the WD, between once every 12000 yrs, such that the WD can cool, to once every 300 yrs, such that it cannot. The system is further complicated by the secular  
drift in the secondary modulation. Such secondary modulation could explain systems like T Pyxidis. Last, we find that the overall process of mass gain by the WD has an efficiency of just $\sim 9\%$, thus requiring a donor with an initial mass of  $\gtrsim 5~\msun$ MS for an initial 1 $\msun$ WD, if the WD is to reach the Chandrasekhar mass. 
\end{abstract}


\section{Introduction}
The problem of mass transfer between interacting stars in close binary systems has been considered for many years. The first results discussed the wide range of time scales that can arise.
\cite{Morton1960} considered evolutionary effects associated with mass exchange, and found that the exchange may be unstable and  lead to significant mass exchange on the thermal, or the Kelvin-Helmholtz-Ritter\footnote{ \cite{Shaviv2010} has shown that Ritter was actually the first to identify the ``Kelvin-Helmholtz" time-scale and the first to calculate it correctly.}  time scale.  \cite{Paczy1967II} followed the evolution of a close massive binary system and demonstrated that the initial mass loss causes the system to depart from thermal equilibrium, while subsequent mass transfer will take place on the evolutionary time of the donor, once it becomes less massive. On the other hand, by applying the methodology of \cite{Morton1960} to stars with convective envelopes, \cite{Paczy1972} have shown that a considerable amount of mass could be transferred on a dynamical time scale, i.e., much faster than radiative envelopes.

As time progressed, it was evident that more complications can arise. \cite{Paczy1967I} studied the secular evolution of massive semi-detached systems and concluded from comparison to observations that about half of the matter transferred from the primary should have been ejected. \cite{Alme1976} studied the initial phase of Roche lobe overflow and found that radiation-driven density waves can induce significant transient mass transfer if the companion is a supergiant. 

\cite{Kippenhahn1977} have shown that the star accreting mass during mass exchange in a binary systems can increase its volume. The authors estimated the effect numerically assuming constant accretion rates. A similar calculation was carried out by \cite{Neo1977}. In both cases the effect on the donor was not calculated, even though an effect could be present. 
 
\cite{Webbink1977I,Webbink1977II} discussed the evolution of close low-mass binary systems ($1.50$~M$_{\odot}$ and $0.50$~M$_{\odot}$ MS stars) and found that the mass transfer is not constant because as mass transfer progresses, the conditions of the donor star change as well. 

\cite{Ritter1988} investigated various mechanisms for on/off switching of the mass transfer in cataclysmic binaries. For example, in an attempt to explain hibernation, it was suggested that mass transfer can be switched off after a nova explosion \citep{Shara1986} or by transferring some of the white dwarf spin into orbital angular momentum \citep{Lamb1987}. However, \cite{Ritter1988} concluded that these mechanisms are insufficient to have large effects. 

The question of mass transfer history is extremely important for the understating of the thermonuclear runaways of classical novae. However, until recently, all analyses assumed a white dwarf accreting at a fixed prescribed rate ${\dot m}$ \citep{Kovetz1998,Starrfield2005,Idan2013}. Usually, these simulations only followed one eruption cycle. The first exception were \cite{Prialnik1995} and \cite{Epelstain2007} who followed $10^3$ cycles in which the WD accretes mass, erupts, ejects the accreted mass, and relaxes towards its original state. It was found that after several cycles, the systems reach a limit cycle such that the initial conditions are forgotten. 

\cite{Idan2013} have shown that there is slow secular change accompanying the eruption cycles due to the accumulation of small changes in the accreting star, and that this can lead to giant flashes. The authors simulated a $1~\msun$ WD accreting at a rate of $10^{-6}~\msun$yr$^{-1}$, and discovered that all the mass accreted up to the 4153$^\mathrm{th}$ nuclear flash was ejected in 3 giant flashes. 

However, the mass transfer rate is expected to be variable. On long time scales, it is secular changes in the donor star which govern the mass transfer rate. For example, \cite{Hameury1991} discussed the evolution of cataclysmic variables and the importance of stellar modeling and suggested intermittent accretion periods.  On shorter time scales, \cite{Kolb1990} demonstrated that the mass transfer rate is a sensitive function of the degree of overflow or underflow of the Roche lobe. Thus, any short term variations in the donor's radius will translate into a variable transfer rate. In the present work we calculate the long term secular evolution of the coupled binary system, but show that variations also exist on time scales shorter than the thermal relaxation time of the donor because of variations in the Roche Lobe overflow.   

Having a variable mass transfer rate can have very interesting repercussions. Many simulations of the progenitors of Type Ia supernovae in singly degenerate models are based on  the calculation of single flashes with a given constant accretion rate. 
However, typical mass accretion efficiencies are less than 10\%, if any net mass is accumulated. Thus, to get a 1 $\msun$ WD increase its mass to the Chandrasekhar limit would require transferring typically 4 $\msun$ of material, over which the characteristics of the donor, and therefore of the accretion, will significantly change. 

We present here the results of a new code that follows the stellar  evolution of the two stars simultaneously on two separated CPU's (or computers) while continuously communicating the relevant parameters describing any mass transfer. These include the accretion rate, the instantaneous composition of the accreted matter and the associated change in orbital parameters, which are then immediately incorporated into the evolution of each star.

The structure of the paper is as follows.
We begin in \S\ref{sec:separation} by discussing the secular evolution of the separation between the two stars. This is necessary for the full treatment of the binary evolution, though we note that the effects discussed in this paper take place on shorter time scales. General aspects of the code we developed are discussed in \S\ref{sec:code}, while the algorithm for mass transfer is detailed in \S\ref{sec:mass-loss}. The results are then presented and discussed in \S\ref{sec:results}. We conclude with a summary in \S\ref{sec:summary}. 

\section{The stellar evolution code}
\label{sec:code}
The stellar evolution code is based on the {\sc evolve} code developed by \cite{Rakavy1967}, which underwent major additions and revisions \citep{Kovetz1970, Shaviv1976I,Shaviv1976II, Godon1989,Kovetz1984,Kovetz1994}. These include updated equation of state, diffusion of isotopes, opacity and hydrodynamics. The implemented version of the nuclear reactions contain 40 species. A unique feature of the code is its separation between solution of the hydrostatics at a fixed entropy density distribution in the star, and solution of the heat equation at  a fixed radial distribution, which allows for fast convergence when rapid adiabatic changes take place. 

The code also solves the hydrodynamic equations whenever the kinetic energy calculated in the hydrostatic phase exceeds a prescribed value. It also contains an algorithm for spherical mass accretion and mass loss in a form of an optically thick and thin winds or an explosion. If the physical conditions allow for the existence of an optically thick wind,  it is assumed that such a wind exists and the ensued mass loss is calculated.  A special routine controls the mass division so as to verify a prescribed accuracy and usually less than 8024 mass shells are considered. This is very important in long calculations which spread over many time steps, and in particular, to describe the secular evolution of the system.  

For the present work, the code has undergone several modifications to allow the simulation of a binary evolution.

\subsection{Simulating two stars}

The novel idea presented in this work is to separately evolve each star on a dedicated CPU while regularly communicating the instantaneous state of each star (in particular the mass transfer rate) to the other one through a dedicated ``gate", allowing both stars to adjust themselves according to the evolving interaction between them. Thus, a single computer with a dual core CPU can run two separate processes with each one simulating the evolution of a single star. 
The idea can be easily extended to systems with more than two interacting stars.  

Since the two stars may evolve at a completely different simulation rate, it is imperative to verify that the clocks run at the same rate for both stars. The clocks are therefore sampled and compared at prescribed intervals. Frequently, one process/star has to wait for the other. The idle CPU is then free for other tasks. Under special conditions however, as is the case with an explosion or sudden mass transfer onto one star, the fixed time intervals at which data is transferred, may decrease significantly. When no interaction between the stars takes place, the time intervals appropriately increase.

\subsection{The evolution of the binary separation}
\label{sec:separation}
The second modification is to evolve the binary orbit. The physics here is straightforward and can be found in text books, e.g. \cite{AccretionBook}. We bring it here for completeness of the description of the code modification. 

Kepler's third law implies that 
\begin{equation}
\Omega^2={GM \over s^3},
\end{equation}
with $s$ being the separation between the binary stars, having a total mass $M=M_d+M_{WD}$. We denote by $M_d$ the donor star, which fills its Roche Lobe during the evolution. We also assume in this simulation that the companion is already a white dwarf of mass $M_{WD}$. However, binary evolution starting from pre-main sequence has been carried out with the code without any problem.

The total angular momentum of the system is given by:
\begin{eqnarray}
J_{tot}&=&J_{orb}+J_{d,spin}+J_{WD,spin} \nonumber \\
&=& \mu s^2 \Omega + k_d M_dR_d^2\omega_d + k_{WD} M_{WD} R_{WD}^2\omega_{WD}.
\end{eqnarray}
Here $\mu \equiv M_dM_{WD}/M$ is the reduced mass while $\omega_d$ and $\omega_{WD}$ are the corresponding spin angular velocities. Also, $k_d$ and $k_{WD}$ are dimensionless numbers which depend on the distribution of angular momentum inside the corresponding stars, and defined as 
\begin{equation}
k_i={1\over M_i R_i^2 \omega_i}\int dm r_i(m)^2 \omega_i(m),
\end{equation}
with $i$ standing for either the donor star or WD. The $k_i$'s are directly related to the radii of gyration for stars when there is no differential rotation.
 In the present calculation we assume that the stars are phase locked and do not rotate, hence, $k_d=k_{WD}=0$.

The energy loss by gravitational radiation is \citep[e.g.,][]{Landau1975}:
\begin{equation}
-{dE \over dt}={32G\mu^2\omega^6s^4 \over 5c^5},
\end{equation}
where $E$ is the mechanical energy of the system. The ensuing decrease in separation is
\begin{equation}
{\dot s}={2s^2 \over GM_dM_{WD}} {dE\over dt}=-{64 G^3M_dM_{WD}M\over 5c^5s^4}. 
\end{equation}
For a circular orbit we have ${\dot E}={\dot J}\Omega$. From the above we obtain:
\begin{equation}
{\dot J}_{grav}=-{32 G^{7/2}M_dM_{WD}M^{1/2}\over 5c^5s^{7/2}}.
\end{equation}

The total angular momentum loss $\dot{J}_{sys}$, which includes the gravitational one as well as from other mechanisms (such as loss by wind), will cause the orbit to evolve. A logarithmic derivative of the orbital angular momentum gives
 \begin{equation}
{\dot {s(t)} \over 2s(t)} ={\dot {J}_{sys} \over J_{orb}} - {\dot{M}_{WD}\over M_{WD}}
\left[1-q(t)-M_{WD}{j_d - j_{WD} \over J_{orb}}\right].
\label{eq:separ}
\end{equation}
where $q=M_{d}/M_{WD}$ and $j_i \equiv J_i/M_i$ are the specific angular momenta.  

We will generally assume in this work that the two stars are synchronized and that the orbit is circular. It is a valid assumption since tidal synchronization and circularization is very rapid in close interacting binaries.  Thus, even though the transfer of mass between the binaries tends to destroy the synchronization, tidal processes will be very quick. Other effects such as magnetic breaking, are neglected as well.    

In our case, most of the change in the separation between the two stars is due to (a) mass transfer between the donor and the white dwarf and (b) the mass loss from the system when the white dwarf erupts and ejects mass. These lead to timescales which are significantly shorter than the gravitational one but longer that the tidal processes.

\subsection{Estimate of mass loss }
\label{sec:mass-loss}
When the donor expands to fill the Roche lobe, mass transfer begins. We assume that 
 the flow through the nozzle between the two lobes is in a hydrodynamic steady state, thus, the flow reaches the speed of sound at the narrowest point of the nozzle.  The rate of mass loss through the $L_1$ nozzle is therefore given by
\begin{equation}
{\dot m}=\rho_{L}v_{sound}Q,
\end{equation}
where $Q$ is the effective cross-section area of the narrowest point of the nozzle, $v_{sound}$ is the isothermal speed of sound and $\rho_L$ the density at the Lagrange point. The above expression can be approximately written as:
\begin{equation}
{\dot m}={1 \over e}Qv_{sound}(T_{ph})\rho_{ph}\exp\left(-{ (R_{*}-R_L)^2 \over H^2}\right),
\end{equation}
where $R_{*}$ is the equivalent radius of the donor star, $R_L$ is the equivalent radius of the Roche lobe, and $H$ is the atmospheric density scale height. 

One of the first treatments of mass transfer through the Roche lobe is by \cite{Meyer1983}, who estimated the effective cross-section to be
\begin{equation}
Q=\pi 2  R_{gas}T_{ph} { s^3\over  G(M_{WD}+M_d)}k,
\end{equation}
where
\begin{equation}
k=\left({ g(M_{WD},M_d)-1 \over g(M_{WD},M_d)} \right)^{1/2}$$
\end{equation}
and 
\begin{equation}
g(M_{WD},M_d)={M_{WD} \over x_{L_1}^3} +{M_d\over (1-x_{L_1})^3}.
\end{equation}
Here $s$ is the separation between the two stars and $ x_{L_1}$ is the distance of $L_1$from the primary in units of $s$. 

\cite{Eggleton1983} provides an estimate for the the Roche radius:
\begin{equation}
R_{L}(t)=s(t){0.49 q^{2/3}\over 0.6q^{2/3} + \ln(1+q^{1/3})},
\end{equation}
which can be evaluated using the instantaneous values of $M_d(t),$ $M_{WD}(t)$ and $s(t)$.
 The latter is given by eq. \ref{eq:separ} under mass transfer between the stars and mass loss from the system.

Another prescription for the mass transfer was given by \cite{Kolb1990}. We implement it in the present version of the code:
\begin{equation}
\label{eq:Kolbmdot}
\dot{m}=\dot {m}_0 \exp \left(- { \Phi_L-\Phi_{ph} \over  \left( R_{gas} T_{ph} / \mu_{ph} \right) }\right),
\end{equation}
where 
\begin{equation}
\dot {m}_0={2 \pi \over \sqrt{e}}F_1(q) {R_{L}^3\over GM_d }\left( R_{gas}T_{ph} \over \mu_{ph} \right)^{3/2}
\rho_{ph}, 
\end{equation}
\begin{equation}
F_1(q)={q\left(s / R_{L}(q) \right)^3  \over \sqrt{g(q)[g(q)-1-q]}},
\end{equation}
and 
\begin{equation}
g(q)=g(M_{d}/M_{WD},1).
\end{equation}
 Here $\rho_{ph},\Phi_{ph} $ and $\mu_{ph}$ are the density, the Roche potential and the mean molecular weight respectively in the photosphere of the donor star.

\begin{figure*}
\includegraphics[width=\figsize]{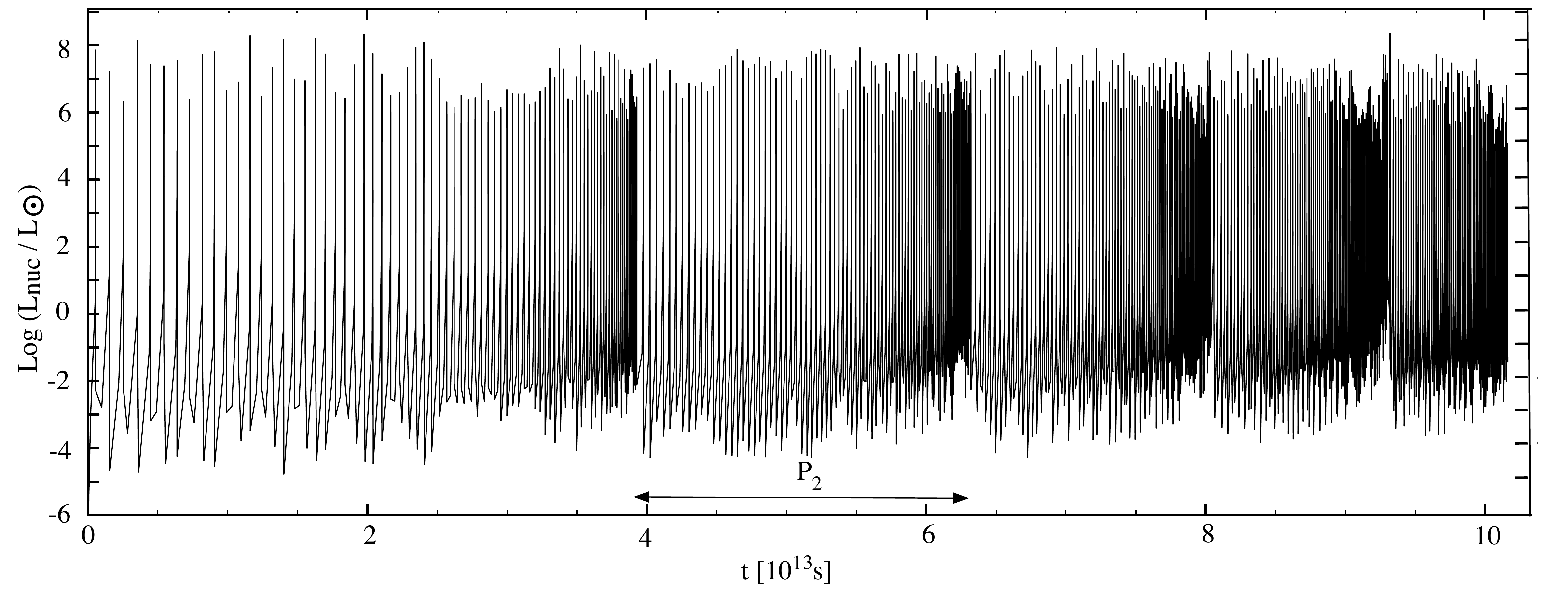}
\caption{\sl   The temporal evolution of the nuclear burning rate $L_{nuc}(t)$, demonstrating the episodic flashes. $t=0$ is defined as the moment mass transfer begins. Also shown is the time period $P_2 \sim \tau_{KHR}$ modulating the flash repetition rate. }
\label{fig:fig-1}
\end{figure*}

\begin{figure*}  
\includegraphics[width=\figsize]{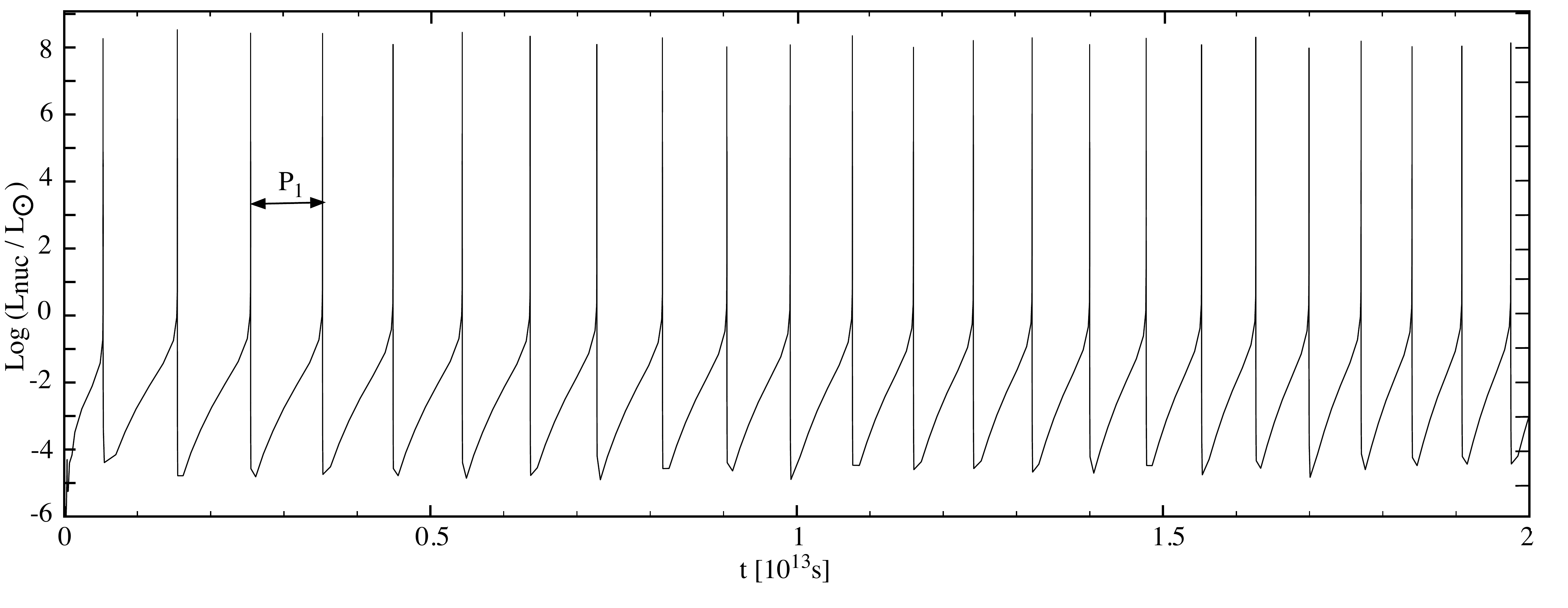}
\caption{\sl   The same as fig.\ \ref{fig:fig-1} with the time axis expanded, showing almost regular periodic flashing.  However, careful inspection reveals that the time between subsequent peaks, $P_1,$ actually decreases as a function of time in this part of the evolution. }
\label{fig:fig-2}
\end{figure*}

The potential difference $\Delta \Phi=\Phi_L - \Phi_{ph}$ is then expanded in the vicinity of $L_1$ in terms of the volume equivalent radius yielding the result:
\begin{equation}
\delta \Phi\approx {R_{L}-R_{ph} \over H_p/\gamma(q)}\left({R_{gas}T_{ph}\over \mu_{ph}} \right) ,
\end{equation}
where $\gamma(q) $ corrects for non-sphericity of the Roche equipotential surface and $H_p$ is the pressure scale height.

This approximation becomes poor  for large $\Delta \Phi$, i.e., for large accretion rates. In such a case,  $\Delta \Phi$ can be evaluated using tables computed by  \cite{Mochnacki1984}.

The condition for the validity of eq.\ \ref{eq:Kolbmdot} is that $\Delta \equiv (R_{ph} - R_{L})/H_p \ll 1$, where  $R_{ph}$ is the photospheric radius \citep{Pastetter1989}. As noted by \cite{Kolb1990}, the mass transfer is very sensitive to $(R_{ph} - R_{L})/H_p$ and usually 
\begin{equation}
(R_{ph} - R_{L}) \approx \left(10^{-5}~\mathrm{to}~10^{-3}\right)R_{*}.
\end{equation}

Other prescriptions for $\dot{m}$ can be found in the literature, for example, \cite{Tout1997} assumed a power law of the form $\dot{m} \propto \left[\left(R_{*}-R_{Roche} \right) / R_{*} \right]^3$, while \cite{Webbink1977I} assumed a linear relation of the form $\dot{m} \propto \left( R_{*}-R_{L} \right) / R_{*} $. Other expressions for the mass transfer can be found in  \cite{Imshennik2002,Kolb2001} and \cite{Hameury1991}. 


\section{Results}
\label{sec:results}
We present the evolution of a close binary system composed of a $1.25$~M$_{\odot}$ MS star with an initial separation of $s(t=0)=69.254$~R$_{\odot}$ from a $1.00$~M$_{\odot}$ WD. The WD is composed of $50\%$  $^{12}{\rm C}$ and $50\%$ $^{16}{\rm O}$ (by mass). The initial composition of the MS star is solar, that is, $X=0.7, Y=0.27, X(^{12}{\rm C})=0.015$ and $X(^{16}{\rm O})=0.015$. The binary evolution without any mass transfer is integrated until the moment the MS expands to fill the Roche lobe such that mass transfer begins onto the WD. This moment is defined as $t=0$. 

\subsection{The accretion rate and the period between flashes}
The temporal evolution of the nuclear luminosity is depicted in figs.\ \ref{fig:fig-1} and \ref{fig:fig-2}. Each spike corresponds to a thermonuclear runaway flash. The most prominent  feature is that the interval between flashes, $P_1$, is itself modulated---the interval appears to exhibit a decreasing sawtooth behavior with a period $P_2$. Moreover, the modulation period $P_2$ is itself slowly decreasing as well.


\begin{figure*}  
\includegraphics[width=\figsize]{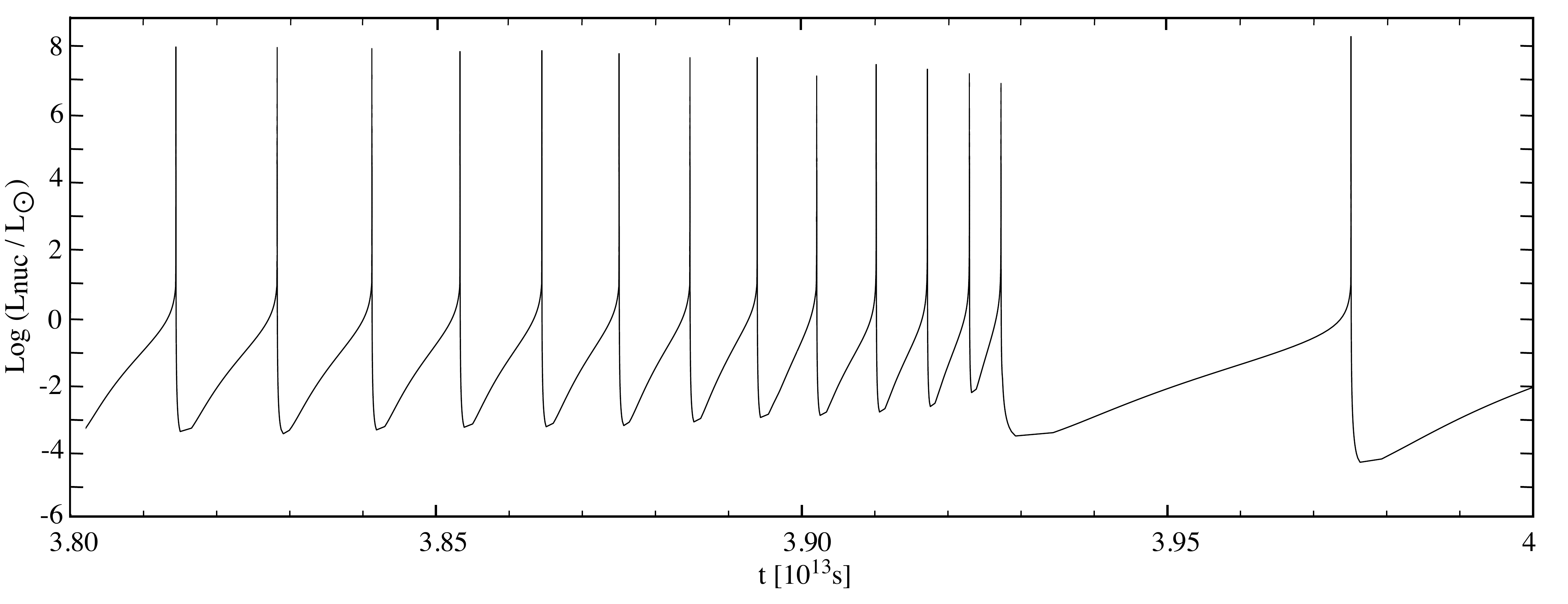}
\caption{\sl   The expansion of the time axis around the flash ``convergence" point. The nuclear luminosity peaks reach a minimal repetition period (maximal mass accretion rate, for the largest radius of the donor star). Subsequently $\dot {m}$ decreases to its minimal value and a new sequence begins.  }
\label{fig:fig-3}
\end{figure*}

\begin{figure*}  
\includegraphics[width=\figsize]{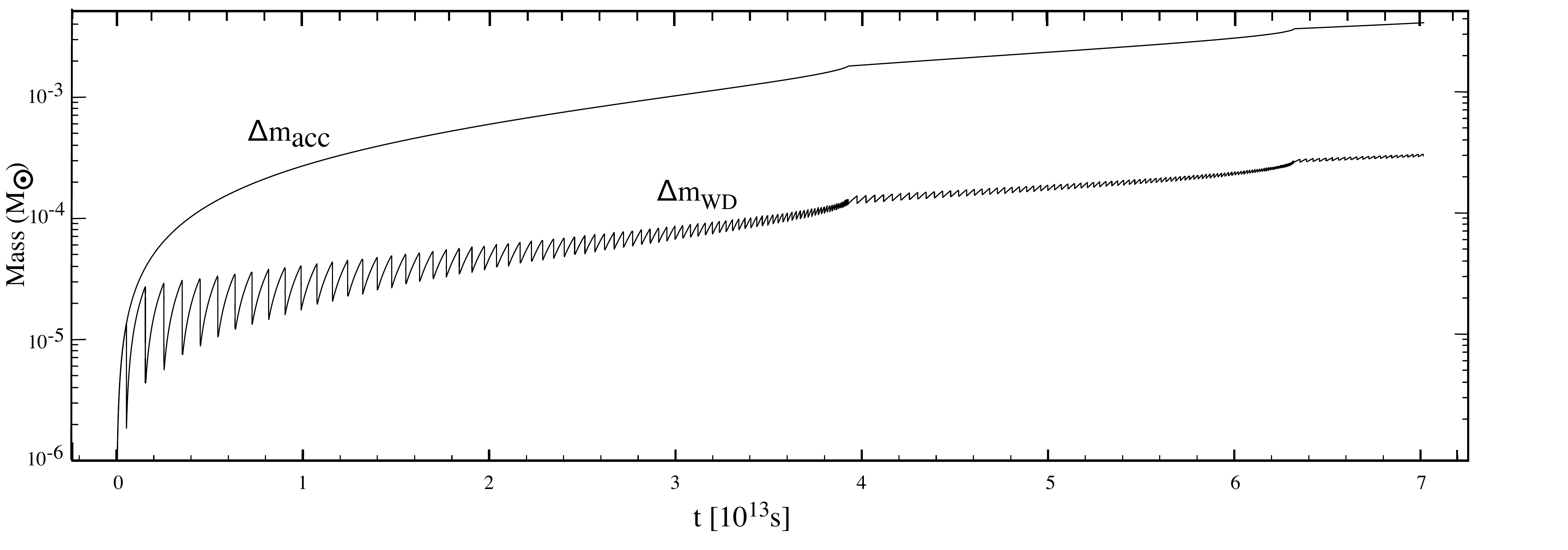}
\caption{\sl   The accumulated mass as a function of time. The fast sawtooth behavior is due to the mass gained and lost during the flashes. The change in the average slope demonstrates that the mass accretion averaged over a few cycles can vary due to the secondary modulation, but since the ratio between $\Delta m_{WD}$ and $\Delta m_\mathrm{acc}$ remains roughly constant (after $t \approx 2 \times 10^{13}$s), the long term efficiency does not change by much.}
\label{fig:fig-4}
\end{figure*}

For the specific system simulated, we find that $P_1$ varies between 12000 yrs to 300 yrs. $P_2$ decreased from about $1.3\times10^6$~yr to $0.2 \times10^6$~yr, but it may continue to decrease (the simulation required too much CPU to run significantly further).

Fig.\ \ref{fig:fig-3} depicts the nuclear burning rate around the time when the interval between flashes suddenly increases again. One can observe that the convergence of the flashes is accompanied by a rise in the minimal value of the nuclear luminosity. However, immediately after the interval increases, the system appears to have sufficient time to be able to relax back to the original pace of nuclear flashes and to cool down such that the minimal inter-flash luminosity decreases again. 

Since the interval between flashes depends on the accretion rate, the secondary sawtooth modulation is a consequence of slow oscillations in the radius of the donor star that govern the mass accretion rate. The oscillations arise because the envelope of the donor is kept out of thermal equilibrium due to the time dependent mass loss. Since previous codes could not describe thermal oscillations that are coupled to the stellar radius and mass loss, this secondary modulation could not have been seen previously.

There are several interesting implications of this result. First, this change can hardly be observed in classical novae because the period is very long ($P_{1} \approx 10^5$~yr, cf.\ \citealt{Bath1978}). However, it could potentially explain the variable recurrence rate in recurrent novae (namely those novae with a period $P_{1}$ which is of order the human observation time of several decades). The recurrent novae T Pyxidis is a typical example (cf.\ \citealt{Godon2014}). This nova had eruptions with maximal apparent magnitude of about 7.0 in the years: 1890, 1902, 1920, 1967 and again after 44 years, in 2011. The nova is recurrent, namely the average period between flashes $P_{1}$ is of the order of 20 years but far from being constant or regular.  

Second, measuring ${\dot m}(t)$ does not necessarily reflect the average mass loss rate because of both the secondary modulation and the long term secular evolution. This implies that it is impossible to link an observationally determined ${\dot m}(t)$ to an evolutionary inferred ${\dot m}(t)$.

\begin{figure} 
\centerline{\includegraphics[width=0.45\textwidth]{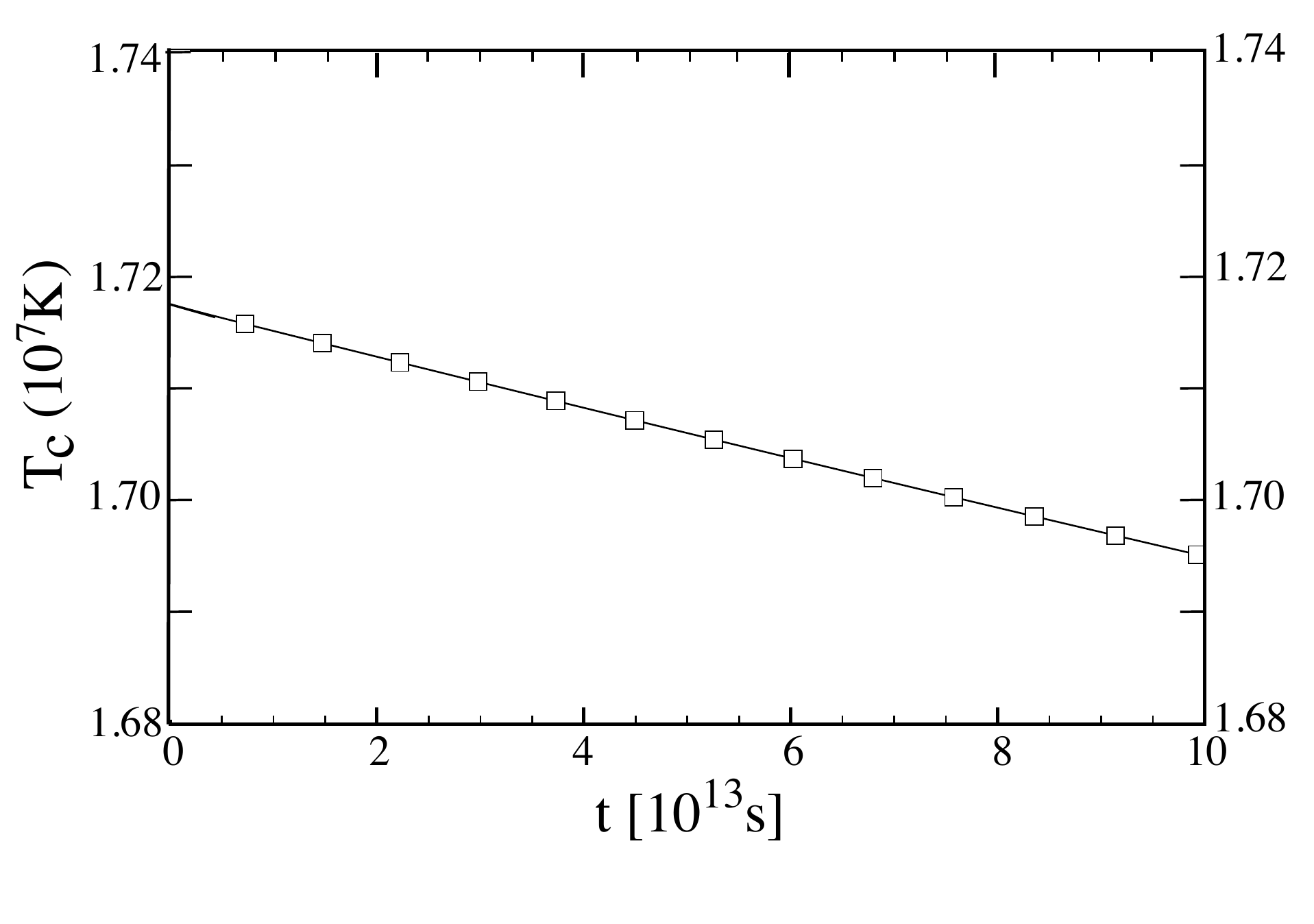}}
\vskip -6mm
\caption{\sl   The central temperature as a function of time.  The core cools, indicating that both the accretion and the adiabatic heating of the WD due to its mass increase are insufficient to heat it.  }
\label{fig:fig-5}
\end{figure}

\subsection{The mass gained by the WD}
The second important aspect of the binary evolution is the mass budget. The evolution of the accumulated mass on the WD is shown in fig.\ \ref{fig:fig-4}. Two main features are apparent.

First, over the flash repetition interval $P_1$, the mass increases during the quiescent phase and then decreases following mass ejection in the nova eruptions. This process is responsible for the fluctuations in the curve.

Second, the ratio between the mass gained and the mass lost in the flash cycles is almost constant, at least after about 20 or so flashes that allowing the WD to reach some sort of equilibrium. Interestingly, the percent retained in the flashes remains roughly the same at the beginning of the secondary modulation cycle when the interval between flashes is long (e.g., at $t = 4 \times 10^{13}$~s) and at the end of the secondary modulation, when the flash interval is short (e.g., at $t = 6 \times 10^{13}$~s). Thus, the ratio between the accumulated mass on the WD $\Delta M_{WD}$ and the total mass accreted mass onto the WD, $\Delta m_\mathrm{acc}$ remains roughly constant over long durations (compare $\Delta m_\mathrm{WD}$ to $\Delta m_\mathrm{acc}$ at $t = 2 \times 10^{13}$~s and at $t = 7 \times 10^{13}$~s). It could however evolve even slower than this computation can detect, which may happen as the average conditions on the WD slowly change. 

As mentioned above, one cannot accurately infer the long term $\left< {\dot m} \right>$ from the instantaneous $\dot{m}$. However, the mass retention fraction over a given cycle does not appear to vary much from the long term average. 

The total mass lost by the donor at $t = 10^{14}$s is   $0.0085$~M$_{\odot}$.
The net mass gained by the WD at the same time is only $0.0008$~M$_{\odot}$. By comparing the two we find that the efficiency in mass accretion $\eta$ is only about $0.09$. 
 
To get a very crude estimate for the initial donor mass required to reach the Chandrasekhar limit, we assume that this mass retention fraction remains without anything dramatic taking place. This itself is questionable given that \cite{Idan2013} found that giant He-flash eruptions can take place and expel the accumulated mass every several thousand hydrogen flashes. If we start with a WD of $1$~M$_{\odot}$ and the system can transfer the {\it entire} donor with the above found efficiency, then the initial donor mass should satisfy
\begin{equation}
M_{d,0} \gtrsim {(M_\mathrm{Ch}-M_{WD,0}) \over \eta} \approx 4.5~\mathrm{M}_{\odot}.
\end{equation}
In reality, the initial mass should be even more massive because the mass transfer will cease long before the entire star is transferred. Lower initial masses are possible only if the mass retention fraction is larger as the WD increases its mass, which we do not know.

In this particular case, the donor has a mass of $1.25$~M$_{\odot}$ and therefore it cannot transfer a significant amount of mass to the WD, though the latter is increasing its mass. This is in conflict with \cite{Godon2014} who claim that T pyx could gain mass only if the mass of the WD is less than $0.9$~M$_{\odot}$, as well as with other recent studies arguing that T pyx loses mass in the eruptions \citep{Selvelli2008,Patterson2013,Nelson2012}. 

\subsection{Secular changes in thermodynamic properties of the WD}

We first remark that the WD responds to each flash and increases its central density as the mass increases. The extra mass of the order of $10^{-7}$~M$_{\odot}$ per flash cycle yields a noticeable change in the central density, demonstrating the calculation accuracy. However, from fig.\ ~\ref{fig:fig-5}, which depicts the secular change in the central temperature, it is evident that the WD cools steadily during the long evolution. In other words, neither the heat from the nuclear flashes at the surface, nor the adiabatic heating due to mass accumulation can compensate for the cooling of the core. 
\section{Summary}
\label{sec:summary}

In this work we modified a stellar evolution code such that it can run simultaneously twice, to describe two interacting stars in a binary system. We then carried out a simulation describing the co-evolution of a WD and a companion star as it leaves the main sequence and begins mass transfer through Roche lobe overflow. The accretion was followed for several million years (of the system), covering several thousand thermonuclear flashes on the WD. The following conclusions were reached:

\begin{enumerate}

\item Coupling the mass transfer rate to the radius of the donor star introduces an instability in the envelope. As a consequence, the radius and the mass transfer rate oscillate with a period somewhat shorter than the Kelvin-Helmholtz-Ritter time scale. 

\item The oscillation period appeared to have a slow secular evolution, becoming progressively shorter. During about 5 oscillations cycles, the period decreased by a factor of 3. The large CPU requirements precluded us from obtaining the steady state periodicity, if a steady state exists. 

\item Because of the variable mass transfer rate, which varied by about a factor of 40, the flash repetition interval varied between about 300 to 12000 yr. The behavior was that of a decreasing sawtooth---every oscillation cycle, the interval started long, progressive decreased, and then jumped to a long value again.    

\item When the repetition interval was long, the WD could cool down between the flashes. When the repetition interval was short, the WD could not cool down. Thus, a mixed behavior is exhibited.

\item The accreted matter in this numerical example comes from the unprocessed envelope of a MS star. As such, it is hydrogen rich. In the flashes, the hydrogen in converted mostly into helium. However, the layer of helium can at some point become too thick and ignite, as found by \cite{Idan2013}. We do not know whether this system with its altered cooling characteristics will exhibit the same behavior. 

\item The process of mass-accumulation has an efficiency of just $9\%$. Such a low efficiency posses severe constraints to  Type Ia supernova scenarios taking place in singly degenerate systems.
\end{enumerate}

\section*{Acknowledgements}
 It is a great pleasure to thanks Dr. Ilan Shaviv for assisting in making this code possible and for providing the required routines for communication between two running processes. This research was partly supported by ISF grant {\#2019020} and a Pazi grant by the IAEC and PBC. NJS gratefully acknowledges support by the IBM Einstein fellowship.

\def\aj{AJ}%
\def\actaa{Acta Astron.}%
\def\araa{ARA\&A}%
\def\apj{ApJ}%
\def\apjl{ApJ}%
\def\apjs{ApJS}%
\def\ao{Appl.~Opt.}%
\def\apss{Ap\&SS}%
\def\aap{A\&A}%
\def\aapr{A\&A~Rev.}%
\def\aaps{A\&AS}%
\def\azh{AZh}%
\def\baas{BAAS}%
\def\bac{Bull. astr. Inst. Czechosl.}%
\def\caa{Chinese Astron. Astrophys.}%
\def\cjaa{Chinese J. Astron. Astrophys.}%
\def\icarus{Icarus}%
\def\jcap{J. Cosmology Astropart. Phys.}%
\def\jrasc{JRASC}%
\def\mnras{MNRAS}%
\def\memras{MmRAS}%
\def\na{New A}%
\def\nar{New A Rev.}%
\def\pasa{PASA}%
\def\pra{Phys.~Rev.~A}%
\def\prb{Phys.~Rev.~B}%
\def\prc{Phys.~Rev.~C}%
\def\prd{Phys.~Rev.~D}%
\def\pre{Phys.~Rev.~E}%
\def\prl{Phys.~Rev.~Lett.}%
\def\pasp{PASP}%
\def\pasj{PASJ}%
\def\qjras{QJRAS}%
\def\rmxaa{Rev. Mexicana Astron. Astrofis.}%
\def\skytel{S\&T}%
\def\solphys{Sol.~Phys.}%
\def\sovast{Soviet~Ast.}%
\def\ssr{Space~Sci.~Rev.}%
\def\zap{ZAp}%
\def\nat{Nature}%
\def\iaucirc{IAU~Circ.}%
\def\aplett{Astrophys.~Lett.}%
\def\apspr{Astrophys.~Space~Phys.~Res.}%
\def\bain{Bull.~Astron.~Inst.~Netherlands}%
\def\fcp{Fund.~Cosmic~Phys.}%
\def\gca{Geochim.~Cosmochim.~Acta}%
\def\grl{Geophys.~Res.~Lett.}%
\def\jcp{J.~Chem.~Phys.}%
\def\jgr{J.~Geophys.~Res.}%
\def\jqsrt{J.~Quant.~Spec.~Radiat.~Transf.}%
\def\memsai{Mem.~Soc.~Astron.~Italiana}%
\def\nphysa{Nucl.~Phys.~A}%
\def\physrep{Phys.~Rep.}%
\def\physscr{Phys.~Scr}%
\def\planss{Planet.~Space~Sci.}%
\def\procspie{Proc.~SPIE}%
 
\bibliographystyle{mn2e} 
\bibliography{References-for-coupled-evolution}{}

\end{document}